# FRONT END CAMAC CONTROLLER FOR SLAC CONTROL SYSTEM

M. J. Browne, A. E. Gromme, Stanford Linear Accelerator Center, Stanford, CA 94025, USA
E. J. Siskind, NYCB Real-Time Computing, Inc., Lattingtown, NY 11560, USA

Abstract

Most of the devices in the SLAC control system are accessed via interface modules in ~450 CAMAC crates. Low-cost controllers in these crates communicate via a SLAC-proprietary bit-serial protocol with 77 satellite control computers ("micros") within the accelerator complex. A proposed upgrade replaces the existing Multibus-I implementation of the micro hardware with commercial-off-the-shelf ("COTS") personal computers. For increased reliability and ease of maintenance, these micros will move from their current electrically noisy and environmentally challenging sites to the control center's computer room, with only a stand-alone portion of each micro's CAMAC interface remaining in the micro's original location. This paper describes the hardware/software architecture of that intelligent front-end CAMAC controller and the accompanying fiber optic link board that connects it to the PC-based micro's PCI bus. Emphasis is placed on the hardware/software techniques employed to minimize real-time latency for pulse-to-pulse operations that control accelerator timing, acquire data for fast feedback loops, and change device settings to close those loops. The controller provides the sole interface between the COTS computing/networking environment and the existing CAMAC plant. It also supports higher bandwidth commercial byte-serial crate controllers and legacy BITBUS hardware.

## 1 INTRODUCTION

The SLAC control system currently utilizes a central database server connected via a proprietary broadband network to 77 front-end microcomputers or "micros" distributed throughout the accelerator complex. Each micro is equipped with a list-processing interface that uses a proprietary 5-megabit/second bit-serial protocol to access 1-4 strings of CAMAC crate controllers. The total number of CAMAC crates is around 450. Approximately 10% of the micros also have BITBUS masters that control PEP-II magnet power supplies.

The database server-to-micro network connection consists of three 1-megabit/second logical segments, with each segment carried in 6 MHz of bandwidth in one of two closed circuit TV ("CCTV") cables via RF modems. Message transfers via this network are asynchronous with the 360 Hz pulsed operation of the SLAC LINAC; the maximum packet length is far longer than the 2.78 millisecond inter-pulse period. Pulse-to-pulse sequencing of accelerator operation is controlled by a 360 Hz broadcast of 128 bits of "trigger pattern" information from the timing control micro. A second, beam-synchronous network carried in another 6 MHz CCTV cable channel distributes this broadcast. Each receiving micro forwards timing information for the next three beam pulses from within the trigger pattern to accelerator devices via an F(19) operation to all slots in all CAMAC crates containing timed devices. Low-latency point-to-point 2-megabit/second baseband links comprise a third, beam-synchronous network. These collect fast-feedback sensor data into a common micro for each feedback loop, and distribute commands to the micros controlling devices that close each loop.

The circa-1980 micro hardware architecture uses the Multibus-I backplane. (A small number of front-end VME/VXI systems that control PEP-II RF stations and other non-pulsed devices are beyond the scope of this article.) Although the processors in these systems have been upgraded to 80386s or 80486s, the performance and density increases in CPU, memory, I/O bus, and networking technology driven by Moore's Law over the past decade have bypassed these systems because of their reliance on an archaic hardware architecture. In addition, the proximity of the micros to high-power pulsed electrical systems has resulted in large amounts of noise pickup in backplanes and communications links, particularly from kickers and klystron modulators. Finally, the micros' remote locations expose them to extremes in temperature and humidity, as well as to large concentrations of airborne dust/dirt.

Preparations for replacing the micros with low-cost COTS personal computers have been underway for three years. The functions of all three existing front-end networks can be combined into a single modern network based on COTS technology such as switched gigabit Ethernet. Quality-of-service primitives within this COTS network ensure adequate real-time latency for beam-synchronous communications supporting trigger pattern dissemination and fast-feedback loops.

Deployment of COTS micros and networks requires the development of the hardware and software of an interface between the micro's PCI bus and the existing CAMAC (and BITBUS) accelerator hardware.

Reliability and maintainability considerations suggest locating the micros in the controlled environment of the control center's computer room. The interface hardware thus is realized as a stand-alone Front End CAMAC controller ("FECC") that is near the actual CAMAC plant and is connected via optical fibers to a PCI Link ("PCIL") board on the micro's PCI bus.

## 2 REAL-TIME CONSIDERATIONS

In the current architecture, sequences of CAMAC operations required by beam-asynchronous threads of execution in the micro are organized into "packages" that require no more than one millisecond for interface execution. Micro software prevents a new package from being passed to the hardware until processing of the previous package has been completed. Beam-synchronous software activated by 360 Hz trigger interrupts queues multiple packages to the hardware without awaiting the completion of previous packages. The maximum latency with which the synchronous software acquires the CAMAC interface after an interrupt is thus one millisecond out of the 2.78-millisecond inter-pulse period. The synchronous software executes the F(19) pulse-to-pulse timing data broadcast, acquires data for fast-feedback loops, and performs other data acquisition and control functions that must occur on a particular beam pulse in order to properly coordinate with actions in other micros on the same pulse or in the same micro on previous and subsequent pulses. Once fast-feedback loop input data are acquired, they are passed to sequences of loop-processing threads of execution within the same micro via inter-thread mailboxes, and to threads in other micros via the third network. These fast-feedback threads have high priority for use of the CPU in their respective micros, but must still spin-wait for the CAMAC interface to be idle before passing a loop-closing package to the hardware. If a lower priority beam-asynchronous thread is allowed to regain control of a micro's CPU while fast-feedback threads await data from another micro, an additional millisecond of latency may intervene before the CAMAC interface can be used to close the loop if the asynchronous thread requests the execution of its own package.

The FECC design addresses these concerns by including separate pipelines for beam-synchronous and beam-asynchronous operations in both the hardware and software. The processing of beam-synchronous operations take precedence whenever these two classes of operations compete for a common resource, and any ongoing asynchronous operation is preempted with a latency no greater than a few microseconds. This design requirement is imposed on the FECC's CAMAC cycle generation hardware, the PCIL's PCI block transfer hardware, the inter-board link, and all software executing on Analog Devices 40-MHz SHARC digital signal processors that add intelligence to both boards.

## 3 HARDWARE FEATURES

The first-generation hardware featured FECC support for four SLAC serial CAMAC cables, PCIL support for a 33 MHz/32 bit PCI, and separate half-duplex 125 megabit/second inter-board data links for beam-synchronous and beam-asynchronous messages. This design employed the Xilinx XC4000XLA family of field programmable gate arrays ("FPGAs"), with three FPGA parts in the PCIL and two in the FECC. The second-generation FECC2 supports four rings of IEEE-standard type L-2 CAMAC crate controllers and a BITBUS system in addition to the four strings of SLAC CAMAC crate controllers. The matching PCIL2 uses a 66 MHz/64 bit PCI, while the second-generation link uses gigabit Ethernet hardware components to support full-duplex 125 megabyte/second transfers. The PCIL2, employing a single XCV1000 FPGA, is currently being debugged. The FECC2's FPGA logic, now nearing design completion, is intended to reside in an XC2V3000. Both FECCs use CAMAC packaging for power, cooling, and mechanical support, but are unconnected to the CAMAC dataway. A PowerPC processor embedded in a Virtex-II FPGA replaces the SHARC in the planned third-generation PCIL3.

In addition to its on-chip memory complement of 64k 32-bit data words and 40k 48-bit program words, the FECC2's SHARC has 512k words of external zero wait state 32/48-bit data/program memory in six 12-nanosecond 4-megabit asynchronous static RAM chips, expandable to 2048k locations with 16-megabit parts. The PCIL2 and FECC2 each have a single 64-kilobyte boot PROM, and can download additional software from the host micro's memory. Operation of the CAMAC, link, and PCI I/O hardware employ Direct Memory Access ("DMA") operations. The combined I/O transfer rates in both boards exceed the capacity of the SHARC's bus and memory. I/O operations thus access special DMA memories attached to each board's FPGA. These employ 133 MHz synchronous static RAM parts operating with the link's 125 MHz clock. The PCIL2 uses one rank of two 32-bit wide parts to achieve 1000 megabytes/second of bandwidth; the FECC2 has four interleaved 32-bit parts providing 500 megabytes/second of bandwidth. Each current part contains one megabyte of storage, and can be upgraded to future 2-megabyte components. Cycles in these memories are allocated via a time-slicing algorithm. The PCIL2's memory allocates 9/16 of the cycles to the PCI interface, 2/16 to each of the link transmit and receive interfaces, and 3/16 to SHARC access to the memory. The FECC2's allocation is 4/16 to each of link transmit and receive, 4/16 to IEEE-standard CAMAC, 1/16 to SLAC CAMAC, and 3/16 to the SHARC. The FECC2's BITBUS hardware has very low bandwidth and employs programmed I/O transfers from the SHARC rather than DMA hardware.

The PCIL's PCI interface performs block transfers at up to 533 megabytes/second, and has hardware support for accessing host memory buffers in paged virtual memory as well as in a linear address space. There are separate control registers for beam-synchronous and beam-asynchronous operations, with an independent pair of read/write data FIFOs for each class. Beam-synchronous transfers acquire the common cycle-generation logic from an ongoing asynchronous transfer with a latency of a few PCI clock ticks.

The link hardware employs a 16-deep ring buffer of pointers to transmit buffers and a similar ring of receive buffer pointers for each transfer class. Buffer contents are segmented into a stream of 608-byte cells, with each cell carrying a 32-byte hardware header and 320 bytes of message payload. The cell is composed of 32 interleaved Reed-Solomon (19,11) codes blocks, each of which carries one header byte, ten payload bytes, and eight bytes of forward error correction code ("ECC"). A beam-synchronous message acquires access to the link from an asynchronous message cell stream with a maximum latency of one 4.864-microsecond cell length. A cell carrying a trigger pattern in its payload has the highest priority for link access. Fields in the cell header permit hardware implementation of independent flow control for each receive buffer, cell receipt acknowledgement, and receipt timeout followed by a cell retransmission sequence, as well as remote register manipulation. The ECC scheme is tolerant of a noise burst more than a microsecond long in every cell. The link protocol design was funded by an SBIR grant for technology for the Next Linear Collider. However, the cell parameters were tailored to the characteristics of the SLAC LINAC klystron modulator pulse, which is most noisy during rise and fall times that are less than a microsecond long and are separated by a 5-microsecond flattop.

The CAMAC hardware for beam-synchronous and beam-asynchronous operations each comprises separate DMA memory interface units and transfer execution units for every one of the eight (i.e. 4 SLAC + 4 IEEE-standard) serial CAMAC cables. Beam-asynchronous block transfers are interrupted in favor of synchronous operations with a maximum latency of one CAMAC cycle. With typical round-trip cable propagation delays this is approximately 10 microseconds with the SLAC protocol and 6 microseconds in the IEEE-standard system. The IEEE-standard hardware also employs an enhanced block transfer protocol used in Kinetic Systems model 3952 crate controllers. This protocol supports pipelined block transfers at the full CAMAC crate dataway rate of 1 microsecond/word. If desired, the hardware can perform a "recovery cycle" to reinitialize a memory pointer register in a CAMAC module with properly incremented data before resuming an interrupted beam-asynchronous operation.

## 4 SOFTWARE FEATURES

The SHARCs on both boards employ custom real-time kernels to provide interrupt-driven CPU scheduling and serialization of access to shared I/O and memory resources. The FECC's kernel supports the full C execution environment, including separate run-time stacks for each thread of execution. The PCIL per-thread context is somewhat smaller as the kernel only supports assembly language programming. This context excludes both hardware and software stacks as well as a substantial number of processor registers. The kernel's interrupt-driven scheduling latency is 3-5 microseconds, and varies with the amount of hardware stack data that must be transferred to/from software stacks during a FECC scheduling event.

The PCIL software's function is largely limited to support for message transfers between the host micro and the FECC, and for FECC access to read/write buffers in micro memory. Forwarding of a trigger pattern from PCI-loadable PCIL registers to the FECC is accomplished without any aid from PCIL software. PCIL support for emulation of the existing CAMAC interface attempts to minimize the number of PCIL-FECC message exchanges by combining commands with write data, and by merging status with read data.

The FECC's application software consists of assembly code to support message transfers and remote micro buffer access and emulate the functions of the existing CAMAC and BITBUS interfaces (with real-time extensions), plus a much larger volume of C code that replaces the functions of the current micro's beam-synchronous software. Moving the execution platform for the micro's synchronous software to the FECC minimizes CAMAC access latency and reduces the micro's real-time load. Separate C entry points exist for initialization, processing a 360 Hz trigger interrupt, and receiving each class of incoming message.

## 5 ACKNOWLEDGEMENTS

This research was funded by the U.S. Department of Energy {"DOE") under SLAC's operating contract and its subcontracts. Additional DOE funding for PCIL2 development was provided by SBIR grant DE-FG02-98ER82628 to NYCB Real-Time Computing, Inc.